\documentclass[aps,prd,floats,floatfix,showpacs,twoside,preprintnumbers,superscriptaddress,nofootinbib]{revtex4}
\usepackage[dvips]{graphicx,pstricks}
\usepackage{amsmath}
\usepackage{amssymb}
\usepackage{epsfig}
\usepackage{color}

\newcommand{\be}{\begin{equation}}
\newcommand{\ee}{\end{equation}}
\newcommand{\ba}{\begin{eqnarray}}
\newcommand{\ea}{\end{eqnarray}}

\def\roughly#1{\mathrel{\raise.3ex\hbox{$#1$\kern-.75em%
\lower1ex\hbox{$\sim$}}}}

\def\gsim{\roughly>}

\def\slashchar#1{\setbox0=\hbox{$#1$}  
   \dimen0=\wd0     
   \setbox1=\hbox{/} \dimen1=\wd1  
   \ifdim\dimen0>\dimen1   
      \rlap{\hbox to \dimen0{\hfil/\hfil}} 
      #1     
   \else     
      \rlap{\hbox to \dimen1{\hfil$#1$\hfil}} 
      /      
   \fi}

\makeatletter
\def\overbracket#1{\mathop{\vbox{\ialign{##\crcr\noalign{\kern3\p@}
\downbracketfill\crcr\noalign{\kern3\p@\nointerlineskip}
$\hfil\displaystyle{#1}\hfil$\crcr}}}\limits}
\def\underbracket#1{\mathop{\vtop{\ialign{##\crcr
$\hfil\displaystyle{#1}\hfil$\crcr\noalign{\kern3\p@\nointerlineskip}
\upbracketfill\crcr\noalign{\kern3\p@}}}}\limits}
\def\upbracketfill{$\m@th\makesm@sh{\llap{\vrule\@height3\p@\@width.7\p@}}%
\leaders\vrule\@height.7\p@\hfill
\makesm@sh{\rlap{\vrule\@height3\p@\@width.7\p@}}$}
\def\downbracketfill{$\m@th
\makesm@sh{\llap{\vrule\@height.7\p@\@depth2.3\p@\@width.7\p@}}%
\leaders\vrule\@height.7\p@\hfill
\makesm@sh{\rlap{\vrule\@height.7\p@\@depth2.3\p@\@width.7\p@}}$}
\makeatother

\begin{document}
\date{\today}


\title{Physical interpretation of the dressed Polyakov loop in the 
Nambu--Jona-Lasinio model}

\author{S.~Beni\' c}
\affiliation{Physics Department, Faculty of Science, 
University of Zagreb, 
Zagreb 10000, Croatia}

\begin{abstract}
We investigate the rapid rise of 
the dressed Polyakov loop in the Nambu--Jona-Lasinio
(NJL) model as a function of temperature.
In QCD such a behaviour is interpreted as a confinement-
deconfinement phase transition.
However, we demonstrate that in the NJL model this is
simply a remnant of the chiral transition. 
 
\end{abstract}
\maketitle

\section{Introduction}

The Polyakov loop \cite{Polyakov:1978vu} is a particular representation of a static
source of color, propagating in compact Euclidean time $\tau$.
When the source has a finite mass, the path in Euclidean space need
not be any more a straight line in $\tau$ - a possibility to deviate
in space directions goes as $\propto 1/m^{|l|}$, where $m$ is the quark mass, and
$|l|$ is the number of links in a 
loop \cite{Gattringer:2006ci,Synatschke:2007bz,Bilgici:2008qy}.
In this case we say that the Polyakov loop is ``dressed'' 
\cite{Gattringer:2006ci,Synatschke:2007bz,Bilgici:2008qy}.

Because of its transformation with respect to the center 
of the color $SU(N_c)$ 
group of Quantum Chromodynamics (QCD) in the 
static limit $m\to \infty$, it can
be used as an order parameter for 
confinement-deconfinement phase transition,
just like the ordinary Polyakov loop \cite{Bilgici:2008qy}.
The dressed Polyakov loop (dPL) has proven to be a valuable tool
in  continuum studies of QCD, with the 
confinement-deconfinement crossover studied in 
the Dyson-Schwinger 
framework \cite{Fischer:2009wc,Fischer:2009gk,Hopfer:2012qr}.

Model calculations of the dPL were performed in the 
NJL \cite{Mukherjee:2010cp,Flachi:2013iia}
as well as in the P(olyakov)NJL 
model \cite{Kashiwa:2009ki} with magnetic field \cite{Gatto:2010qs}.
In particular, one of the results of 
Ref.~\cite{Mukherjee:2010cp} shows that
even in the NJL model
the dPL shows a rapid rise when proceeding from low to high 
temperatures.
As the NJL model is not 
confining\footnote{A simple reasoning behind this statement is that 
quark propagator in he 
NJL model is just the
tree level fermion propagator, albeit with a 
constituent ($\sim 300$ MeV), 
rather than current quark mass ($\sim 5$ MeV).
Therefore, a positive definite spectral representation 
is available, allowing the excitation of these states.}, it should 
be clear then
that a physical interpretation of the behavior of the dPL
in the NJL model will not be the same as in QCD.

In this paper we consider a natural 
interpretation
of the dPL in the NJL model.
By simple Landau analysis we will understand 
how it comes to be that the dPL
in the NJL model exhibits a rapid change as a 
function of the temperature in the first place.
We will analytically show that the temperature 
at which the change is most
pronounced is, irrespective of the model details, 
the chiral restoration temperature.
With this result we can 
understand that the crossover behavior 
in the dPL calculated in the NJL model 
should be interpreted merely as an imprint of the 
chiral phase transition. 

\section{Nambu--Jona-Lasinio model with twisted boundary conditions}

We work in the chiral limit with $N_f=2$, 
and zero real chemical potential, $\mu=0$.
In order to calculate the dPL one has to 
distort the fermionic boundary conditions, 
by introducing a twisted
angle $\phi$ \footnote{Details are omitted for simplicity, 
interested reader is invited
to consult \cite{Gattringer:2006ci} or \cite{Fischer:2009wc}.}.
Alternatively, one can start from the 
imaginary chemical potential,
so that the thermodynamic potential in 
NJL model is given as \cite{Mukherjee:2010cp}
\be
\Omega = \Omega_\mathrm{cond}+\Omega^\mathrm{kin}_\mathrm{vac}+
\Omega^\mathrm{kin}_\mathrm{th}~,
\label{eq:pot_tot}
\ee
where the condensation potential is $\Omega_{\mathrm{cond}}=\sigma^2/2G$, and
the vacuum and thermal one-loop contributions read
\be
\Omega^\mathrm{kin}_\mathrm{vac} = -d_q\int \frac{d^3 p}{(2\pi)^3} \frac{E}{2}~,
\label{eq:pot_vac}
\ee
\be
\Omega^\mathrm{kin}_\mathrm{th} = -\frac{d_q}{2}T\int \frac{d^3 p}{(2\pi)^3} 
[\log(1+e^{-\beta (E+i\mu_I)})+(\mu_I\to -\mu_I)]~,
\label{eq:pot_therm}
\ee
where $E=\sqrt{\mathbf{p}^2+\sigma^2}$, and $d_q = 2\times 2\times N_f\times N_c$.
Divergence of the vacuum energy (\ref{eq:pot_vac}) 
is regulated by a cutoff $\Lambda$.
The mean field $\sigma$ is obtained by minimizing the 
thermodynamic potential (\ref{eq:pot_tot}).
A nonzero value of the mean field $\sigma$ signals chiral symmetry breaking.
Equivalently, one considers the quark condensate $\langle\bar{q}q\rangle = -\sigma/G$
as an order parameter.

Twisted boundary conditions 
for the fermion field 
$\psi(\mathbf{x},\tau)=e^{i\phi}\psi(\mathbf{x},\tau+\beta)$, 
are equivalent to setting
$\mu_I = T(\phi-\pi)$, with $\phi \in [0,2\pi\rangle$.
The statistically correct fermion degrees of 
freedom are obtained with $\phi=\pi$.
However, it is important to notice that, using $\phi=0$ 
does not make these fields bosons.
Only by altering the overall sign in the vacuum and thermal one-loop 
contributions does one obtain a true Bose
potential.

\subsection{No symmetry restoration at the boundary}

We employ a 
$\sigma/\Lambda\ll 1$, and a $\sigma/T\ll 1$ expansion in 
the vacuum and thermal
parts, respectively.
The relevant expressions are well known in the literature: the vacuum part can be
found in \cite{Yagi:2005yb}, and the thermal part e. g. in \cite{Actor:1986zf}.
To discuss second order chiral phase transition, quadratic 
part has to contain vacuum and thermal fluctuations, while for the quartic part one can
stick just to the vacuum contribution
\be
\Omega(\sigma) \simeq -\frac{1}{2} a(T,\phi)\sigma^2 -
\frac{d_q}{64\pi^2}\log\left(\frac{\sigma^2}{4\Lambda^2}\right) \sigma^4 ~,
\label{eq:pot_lan}
\ee
where
\be
a(T,\phi) = a_0 + \frac{d_q T^2}{2} B_2\left(\frac{\phi}{2\pi}\right) = 
a_0 + \frac{d_q T^2}{8\pi^2}
\left[(\phi-\pi)^2 - \frac{\pi^2}{3}\right]~.
\label{eq:quad}
\ee
and $B_2(x)$ is the second Bernoulli polynomial.
The factor $a_0$ is just the quadratic vacuum contribution
$$a_0 = \frac{1}{G_c}-\frac{1}{G}~,$$
where $G_c \Lambda^2 = 8\pi^2/d_q$.
For fermion boundary conditions $\phi=\pi$, the 
usual role of thermal fluctuations is to flip the sign
of the quadratic term, marking the critical temperature.

In the case of general $\phi$ it is interesting that the
thermal contribution of $a(T,\phi)$ itself can change sign.
This happens at 
$$\phi_\pm = \pi \pm \frac{\pi}{\sqrt{3}}~.$$
Namely, in the region $\phi_-<\phi<\phi_+$, which we 
call {\it fermion-like}, the model
can be subjected to a standard symmetry-breaking-restoration scenario 
provided that the symmetry is broken in the vacuum, i. e. $G>G_c$.
This is the usual case in the NJL model.
However, at {\it boson-like} twisted angles $0\le\phi<\phi_-$, 
$\phi_+ <\phi <2\pi$, the quark condensate
will not respond to arbitrary high thermal excitations.
In other words, the critical temperature obtained 
from the condition $a(T,\phi)=0$
\be
T_c (\phi) = \frac{8\pi^2}{d_q a_0}\frac{1}{\frac{\pi^2}{3}-(\phi-\pi)^2}~,
\ee
diverges at the boundaries.
For convenience we denote $T_\chi = T_c(\pi)$.

Thus, the only way for the mean field $\sigma$ to be zero
at boson-like angles 
is by altering the theory by hand. 
For example, if we choose $a_0<0$, i. e. $G<G_c$ - then we
find ourselves in a weird position where the model 
has a restored phase at low
temperatures, and a broken phase at high temperatures.
The other possibility would be to acknowledge the 
fact that bosonic theories
perfectly well break and restore symmetries as the fermionic ones.
More precisely, if we understand the vacuum 
contribution in (\ref{eq:pot_tot}) 
as a potential term in a classical 
bosonic $Z(2)$ theory, then the thermal 
contribution has to have an {\it ad hoc}
sign change if the thermal fluctuations 
are also understood as coming 
from bosonic fields.
Only then this bosonic theory breaks the 
symmetry at low, and restores it at high
temperatures.

\subsection{Qualitative behavior of the dressed Polyakov loop}

Strictly speaking, the dPL can be defined only when the 
quark mass is non-zero \cite{Bilgici:2008qy}.
Naively, one can still calculate this quantity by 
using its definition \cite{Bilgici:2008qy} as a first 
Fourier mode of the quark condensate at non-trivial twisted angle
\be
\Sigma_1(T) = 
\int_0^{2\pi} \frac{d\phi}{2\pi} \, e^{i\phi} \, \langle \bar{q}q\rangle(T,\phi)~.
\label{eq:dpl}
\ee
We will now use arguments of the previous subsection to 
qualitatively understand that $\Sigma_1(T)$ has to rapidly change 
across the chiral phase boundary. 

First of all, by letting $T\to 0$, the condensate 
does not depend on $\phi$.
This is because the generalized boundary conditions 
modify only the thermal
part (\ref{eq:pot_therm})\footnote{Actually, 
this is tantamount of saying that the Polyakov loop $\Phi$
itself will be zero strictly at $T = 0$ regardless of 
the fact that the 
theory is confining or not.
That is, even if the free energy $F$ of a static quark
is finite, we have that $\Phi = e^{-F/T} =0$ since $T=0$.}.
Therefore, at $T\simeq 0$, we conclude that $\Sigma_1 \simeq 0$.

However, slightly above the chiral restoration 
$T\gsim T_\chi$,
chiral symmetry is first restored in a small region 
around $\phi = \pi$.
This allows for a non-trivial 
Fourier transform (\ref{eq:dpl}), establishing
a non-zero $\Sigma_1$.
Therefore, it appears that as long as chiral symmetry is broken
in the vacuum, i. e. $a_0>0$, the dPL will inevitably display a
significant change proceeding from low ($T\ll T_\chi$) to high
($T\gg T_\chi$) temperatures. 

\section{Divergence of the temperature 
derivative of the dressed Polyakov loop}

\begin{figure}[!t]
\begin{center}
\psfig{file=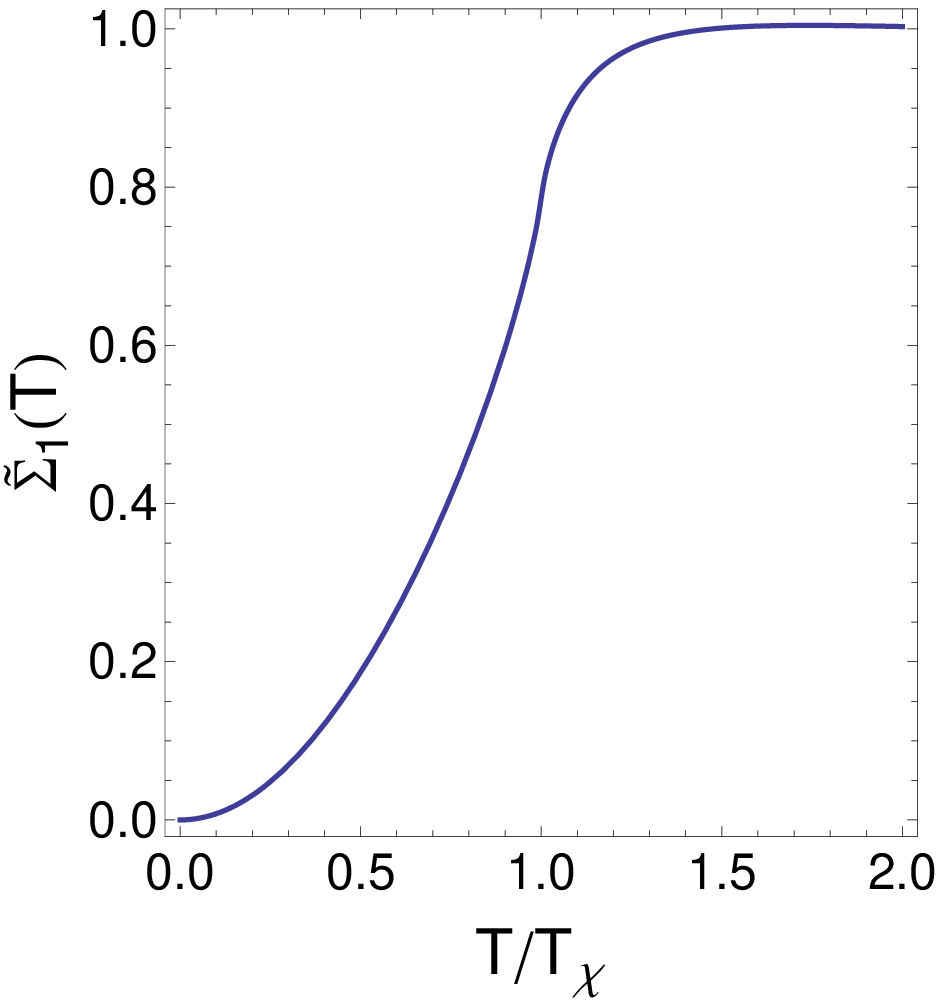,scale=0.5}
\hspace{0.5cm}
\psfig{file=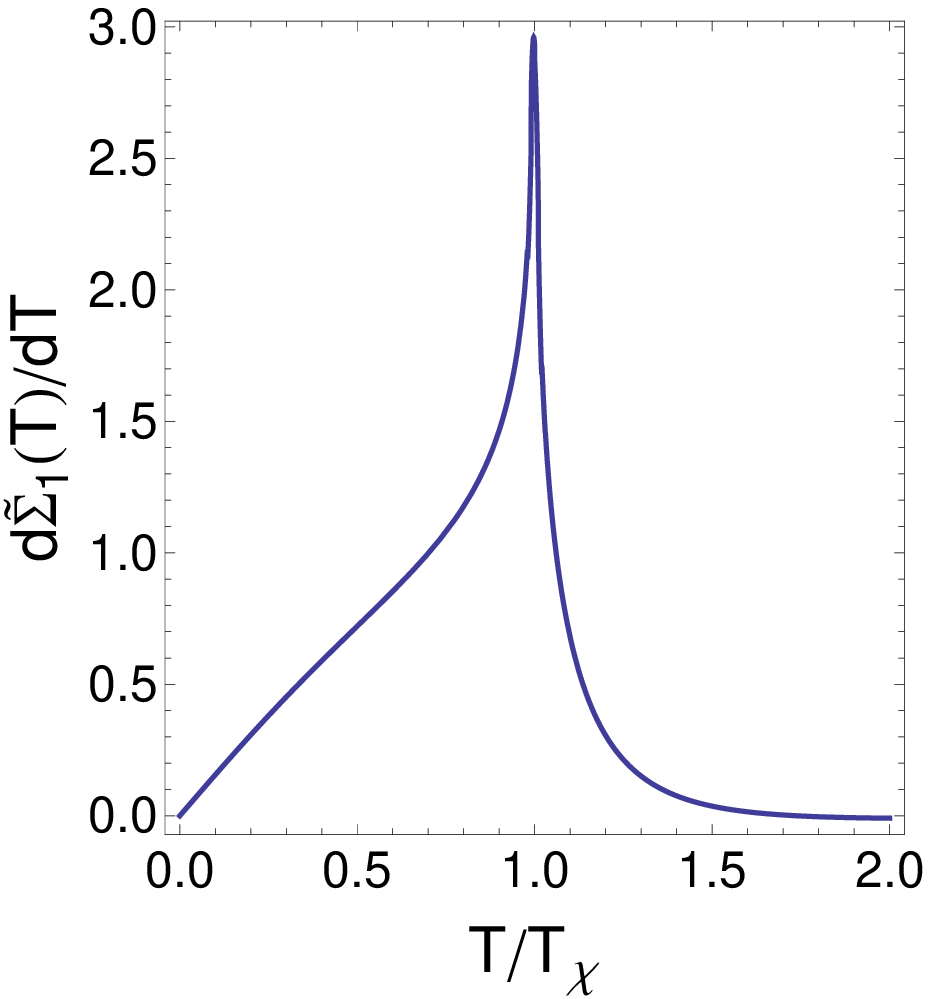,scale=0.5}
\caption{
On the left panel the dPL as a function of temperature 
is shown as calculated from the approximation to
the condensate (\ref{eq:lam}),
while the right panel provides the temperature derivative.
The explicit value is normalized to the high temperature behavior, i. e. $\tilde{\Sigma}_1(T) = \Sigma_1(T)/\Sigma_1(\infty)$.
}
\label{fig:dpl}
\end{center}
\end{figure}
From a general thermodynamical point of view it is known 
that the phase transition leaves 
its mark on all thermodynamic quantities
calculated from the partition function.
For example, a second order chiral phase transition leads to a 
non-analyticity in
the second derivatives of the thermodynamic potential, e. g. 
chiral or thermal susceptibility,
heat capacity, and so on.
Therefore, it is not unreasonable to expect that the 
dPL in the NJL
model should also display similar properties not because the NJL model describes 
confinement as well as the
confinement-deconfinement phase transition, but simply because it does a 
good job in describing the chiral one. 

Let us now look for the temperature where the value of $d\Sigma_1 / dT$
has a maximum.
By acknowledging the fact that $\langle\bar{q}q\rangle$ can be zero at 
temperatures $T>T_\chi$, we have
\be
\Sigma_1(T)=\int_0^{\phi_c(T)} \frac{d\phi}{\pi} \, \cos\phi \, \langle\bar{q}q\rangle(T,\phi)~,
\label{eq:sigc}
\ee
where the upper limit of integration
is depending on the temperature, 
the specific values given by solving $a(T,\phi)=0$ for $\phi$
\be
\phi_c(T) = \pi - 
\frac{\pi}{\sqrt{3}}\left(1-\frac{T_\chi^2}{T^2}\right)^{1/2}~.
\label{eq:phit}
\ee

Using simple algebra, the quantity $d\Sigma_1(T)/dT$ is given as
\be
\frac{d\Sigma_1}{d T} = 
\int_0^{\phi_c(T)} \frac{d\phi}{\pi} \, \cos\phi \,
\frac{\partial\langle\bar{q}q\rangle(T,\phi)}{\partial T}
+\frac{d \phi_c(T)}{d T}
\frac{\partial}{\partial\phi}\left[\frac{1}{\pi}\langle\bar{q}q\rangle(\phi,T) 
\,\cos\phi \right]_{\phi=\phi_c(T)}~.
\label{eq:dsig2}
\ee
Here
\be
\frac{d\phi_c(T)}{d T} = -\frac{\pi}{\sqrt{3}}\, \frac{1}{T} \,
\frac{T_\chi^2}{T^2}\left(1-\frac{T_\chi^2}{T^2}\right)^{-1/2}~.
\label{eq:dphi}
\ee
We now realize that (\ref{eq:dphi}) diverges as
$T\to T_\chi$ from above.
If we naively assume that the first term in (\ref{eq:dsig2}), 
as well as the one multiplying (\ref{eq:dphi}),
is smooth across
the phase transition, then
$d\Sigma_1/dT$ would have a maximum, or -- more 
precisely -- would diverge, 
exactly at $T = T_\chi$. 

In an actual calculation, it turns out that the critical behavior itself is
``one level'' milder.
The solution of the gap equation $\partial \Omega /\partial\sigma = 0$
with the truncated thermodynamic potential (\ref{eq:pot_lan})
is given in terms of the well known 
Lambert $W_{-1}$-function \cite{Corless:1996zz,Yagi:2005yb}.
Thus, the thermal behavior of the condensate for 
general $\phi$ can be approximated
as
\be
\langle\bar{q}q\rangle (T,\phi) \simeq -\frac{2\Lambda}{G} 
\exp\left[-\frac{1}{4}
-\frac{1}{2}W_{-1}\left(-\frac{4\pi^2 e^{1/2} a(T,\phi)}{d_q \Lambda^2}\right)\right]~,
\label{eq:lam}
\ee
which is to be used only in the fermionic-like region.
In the bosonic-like region the mass gap is finite, so the 
$\sigma/\Lambda, \sigma/T\ll 1$ expansion is no longer applicable, 
but we might just approximate the true solution 
in this region with its
vacuum value.
This is simply (\ref{eq:lam}) with a replacement $a(T,\phi) \to a_0$.
Then we can use this in order to integrate (\ref{eq:sigc}).
We use the parameters of Ref. \cite{Grigorian:2006qe}, 
where $G\Lambda^2 = 4.636 $ and
$\Lambda = 602.472 $ MeV.
Fig. \ref{fig:dpl} shows the result, where in the derivative of dPL
instead of the naive divergence, there is a sharp cusp structure.

We stress that a similar cusp behavior
was seen in lattice QCD calculations in the strong coupling 
limit \cite{Fromm:2011kq}.
Whereas the NJL model is non-confining, it is interesting 
that in Ref.~\cite{Fromm:2011kq}
one deals with a completely opposite situation:
thanks to the fact that the system is strongly coupled, deconfinement
does not occur, so the change in the Polyakov loop,
and, in particular, the cusp is indeed interpreted as an imprint of the 
chiral transition; see Figs.~2 and 3 in \cite{Fromm:2011kq}.

\section{Conclusions}

Because the NJL interaction 
dresses the quark with a momentum
independent mass, the singularity structure
of the quark propagator 
is very simple, which is
usually interpreted as a lack of confinement
\footnote{For example, in NJL model 
calculations this lack of confinement 
usually leads to the $\rho$ meson
mass lying above the kinematic threshold for $\bar{q}q$ decay.}.
However, as shown here, and
numerically by Ref.~\cite{Mukherjee:2010cp}, 
calculation of the dPL within the NJL model leads
to a order parameter-like behavior.

The semi-analytic study performed here demonstrated that the change
in the temperature behavior of 
the dressed Polyakov loop in the NJL model 
is entirely dictated by the chiral transition.
Our second result is that the temperature where $d\Sigma_1/dT$ 
diverges
and the chiral transition temperature coincide exactly in the chiral limit.
In Ref.~\cite{Mukherjee:2010cp} the latter result was obtained numerically for a
particular set of parameters.
We have shown that their result is more general, i. e.
independent of the model parameters.

\section*{Acknowledgments}

We thank D.~Horvati\' c for useful comments and
a critical reading of the manuscript.
Helpful discussions with B.~Klajn, O.~Philipsen and W.~Unger are acknowledged.
S. B. is supported by the Ministry of Science, Education and Sports of Croatia
through the Contract No. 119-0982930-1016, and by the CompStar Network.

\end{document}